\documentclass[10pt, a4paper]{article}
\usepackage[usenames, dvipsnames]{xcolor}
\usepackage{lrec2022} 
\usepackage{multibib}
\newcites{languageresource}{Language Resources}
\usepackage{graphicx}
\usepackage{tabularx}
\usepackage{soul}
\usepackage{enumitem}
\usepackage{multirow}
\usepackage[utf8]{inputenc}
\usepackage{csquotes}
\usepackage{booktabs}
\usepackage{subfig}
\usepackage{arydshln}
\usepackage{epstopdf}
\usepackage{hyperref}
\usepackage{xstring}
\usepackage{color}
\usepackage{todonotes}
\usepackage{textcomp}
\usepackage{amsmath,amssymb}
\usepackage{soul}
\usepackage{array}

\setlist{nolistsep}
\usepackage{titlesec}
\titleformat{\section}{\normalfont\large\bfseries\center}{\thesection.}{1em}{}
\titleformat{\subsection}{\normalfont\SmallTitleFont\bfseries\raggedright}{\thesubsection.}{1em}{}
\titleformat{\subsubsection}{\normalfont\normalsize\bfseries\raggedright}{\thesubsubsection.}{1em}{}
\renewcommand\thesection{\arabic{section}}
\renewcommand\thesubsection{\thesection.\arabic{subsection}}
\renewcommand\thesubsubsection{\thesubsection.\arabic{subsubsection}}

\title{FACTOID: A New Dataset for Identifying\\ Misinformation Spreaders and Political Bias}

\name{Flora Sakketou$^{*\dagger}$, Joan Plepi$^{*\dagger}$, Riccardo Cervero$^\ddagger$, Henri-Jacques Geiss $^\diamond$, \\
{\bf \large Paolo Rosso$^\ddagger$ , Lucie Flek$^\dagger$}}

\address{$^\dagger$Conversational AI and Social Analytics (CAISA) Lab\\
Department of Mathematics and Computer Science, University of Marburg, Germany\\ 
$^\ddagger$ Pattern Recognition and Human Language Technology (PRHLT) Research Center\\
Universitat Politècnica de València, Spain\\
$^\diamond$ Department of Computer Science, Technical University of Darmstadt \\
$^*$ These authors contributed equally to this work \\
         \{flora.sakketou, joan.plepi, lucie.flek\}@uni-marburg.de, 
         \{rcerver1@upvnet, prosso@dsic\}.upv.es \\ henri-jacques.geiss@stud.tu-darmstadt.de}

\abstract{
Proactively identifying misinformation spreaders is an important step towards mitigating the impact of fake news on our society. In this paper, we introduce a new contemporary Reddit dataset for fake news spreader analysis, called FACTOID, monitoring political discussions on Reddit since the beginning of 2020. The dataset contains over 4K users with 3.4M Reddit posts, and includes, beyond the users' binary labels, also their fine-grained credibility level (very low to very high) and their political bias strength (extreme right to extreme left). As far as we are aware, this is the first fake news spreader dataset that simultaneously captures both the long-term context of users' historical posts and the interactions between them. To create the first benchmark on our data, we provide methods for identifying misinformation spreaders by utilizing the social connections between the users along with their psycho-linguistic features. We show that the users’ social interactions can, on their own, indicate misinformation spreading, while the psycho-linguistic features are mostly informative in non-neural classification settings.  In a qualitative analysis we observe that detecting affective mental processes correlates negatively with right-biased users, and that the openness to experience factor is lower for those who spread fake news.
 \\ \newline  }

\begin{document}

\maketitleabstract

\section{Introduction}

As the popularity of social media platforms continuously grows, so does the dissemination of online disinformation. Many deep learning systems have been therefore developed to detect false or biased news \cite{10.1145/3395046,DBLP:journals/corr/abs-1905-12616,DBLP:journals/corr/abs-1902-06673,DBLP:journals/corr/abs-1708-01967}. While fake news detection is a big step to mitigate the impact of misinformation on our society \cite{FIGUEIRA2017817,VISENTIN201999}, it is not sufficient, since limiting the diffusion of false information and avoiding its catastrophic effects is extremely challenging, especially once it has been shared on the Web  \cite{doi:10.1080/15205436.2020.1750656,doi:10.1177/1065912920938143}. Research shows that fact corrections frequently fail in reducing people's misconception of the truth, and occasionally they even have a ``backfiring'' effect where people's misconception is reinforced \cite{https://doi.org/10.1111/j.1467-9221.2010.00772.x,10.2307/40587320,2db5d0ff768848249835358b0e8548f5,berinsky_2017}.

It is essential to address this issue at its origin - to efficiently and rapidly identify accounts and users which are likely to propagate posts from the handles of unreliable news sources. While there are numerous datasets focusing on this issue at a post-level, only very few of those allow to approach this matter on a user level, since, in most cases, fake news posts are not associated with their individual authors. 

Moreover, existing datasets designed for identifying misinformation spreaders only include binary labels for the users. However, reality is not black and white, therefore a credibility score associated with each user is more realistic. In addition, since partisan polarization constitutes one of the primary drivers of political fake news sharing \cite{osmundsen_bor_vahlstrup_bechmann_petersen_2021}, it is becoming all the more vital to explore the political bias of users in combination with their misinformation spreading behavior. To the best of our knowledge, there is no existing dataset that combines both of these dimensions on a user level with fine-grained scores.

To this end, we introduce a dataset for distinguishing the authors that have shared news from unreliable sources in the past, from those that share news from reliable sources, covering the posting activity of the users before and after the 2020 US presidential elections. We use the terms \textit{misinformation spreaders} and \textit{real news spreaders}, respectively. Apart from the binary labels, we assign a credibility score to each user based on the factuality of the news sources they shared, and a political bias score based on the level of partisanship of the news sources they share.

Our contributions can be summarized as follows:

\begin{itemize}
    \item We introduce FACTOID\footnote{{https://github.com/caisa-lab/FACTOID-dataset}}: a user-level \textbf{FAC}tuality and p\textbf{O}litical b\textbf{I}as \textbf{D}ataset, that contains a set of 4,150 news-spreading users with 3.3M Reddit posts in discussions on contemporary political topics, covering the time period from January 2020 to April 2021 on individual user level.
    \item Additionally, we provide fine-grained scores about the users' factuality and political bias.
    \item We conduct classification experiments for identifying misinformation spreaders by utilizing the social connections between the users along with their posting history representations and psycho-linguistic features. 
    \item The curated dataset preserves the structure of the threads, facilitating the exploration of the users' social activity by modeling it in a graph. We show that the users’ social interactions can, on their own, indicate misinformation spreading, when used in a graph attention network.
    \item We conduct qualitative analysis of the impact of various psycho-linguistic features, such as affective mental processes and openness to experience.
\end{itemize}

\section{Related Work}
\paragraph{Relevant Datasets. }

User profiling approaches have been investigated for various tasks, such as author profiling \cite{Reddy2016ASO}, bot detection \cite{Cai-2017Beh,Hurtado-2019Bot,Kosmajac-2019Twi}, gender detection \cite{pan19}, among others. However, fake news spreader detection is an under-explored research direction. There are some datasets approaching this matter from different angles, for example, attempts have been made to analyze the users reactions to fake news \cite{DBLP:journals/corr/abs-1805-12032} or to analyze users who debunk fake news \cite{10.1145/3331184.3331248}. However, there are only a few publicly available datasets suitable for the task. 

\newcite{8397048} constructed a dataset by assessing the users' trust level on fake news. More recently, the PAN 2020 competition \cite{Rangel-2020Ove} brought the problem of misinformation spreaders identification to the fore. The dataset of the competition contained 500 users with 100 posts each, for two languages. \newcite{10.1007/978-3-030-51310-8_17} and \newcite{DBLP:journals/peerj-cs/MuA20} created a dataset containing misinformation and real news spreaders by collecting users that posted articles that have been debunked as fake and built their user history based on their previous posts. We draw inspiration from the method of curation of these datasets and use a similar semi-supervised method to obtain a description of the authors and their context. However, the proposed dataset is distinctive in three aspects: it contains fine-grained labels about (a) the users' credibility and (b) political bias, and (c) it preserves the structure of the threads. Additionally, while the aforementioned datasets utilized Twitter as a source, we utilize Reddit which does not have a word limit on the posts, making the task all the more challenging.

\paragraph{Approaches to Spreader Detection.}

Our dataset preserves the structure of the threads, facilitating the exploration of the users' social activity by modeling it in a graph.
The recent advances in graph representation learning \cite{Wu_2021} in various domains provide a promising, under-explored research direction in the context of fake news spreader detection. More specifically, Graph Attention Networks (GAT) \cite{velickovic2018graph} have achieved state-of-the-art-results in various natural language processing tasks \cite{plepi2021perceived,sawhney-etal-2021-suicide,kacupaj-etal-2021-conversational,Ren2020HGATHG}. However, this method has not been explored on user graphs in the context of fake news spreader detection. Research has shown that users tend to interact with like-minded individuals \cite{Bahns17}. Therefore, we wish to leverage this attribute in order to obtain better user representations.

Traditional feature-based user modeling methods analyze the users' linguistic patterns in order to infer psycho-linguistic features \cite{TausczikPennebaker,Girlea2016}. These works extract evidence of mental processes through the Linguistic Inquiry and Word Count (LIWC) software in order to tackle the problem of identifying deceptive authors. Certain psycho-linguistic characteristics are assumed to underlie the vulnerability to fake information, therefore the LIWC tool has often been used to investigate the phenomenon of misinformation from both document-level \cite{Zhou2020FakeNE,PrezRosas2018AutomaticDO} and user-level perspectives \cite{10.1007/978-3-030-51310-8_17,Cervero}. Interestingly, this method has been used in comparison and in conjunction with innovative graph-based architectures \cite{Ren2020HGATHG}. Therefore, we believe that leveraging these psycho-linguistic features and their combination together with the users' social interactions can contribute in order to obtain a strong, competitive baseline. 

\section{Dataset}
\subsection{Terminology}

The term \textit{misinformation} in this paper is used specifically in the context of politics as an umbrella term that covers many aspects: (a) \textit{misinformation}: any news that is false or misleading but is not intended as such, (b) \textit{disinformation}: any false or misleading information that is spread with the specific intent of deception, (c) \textit{hyperpartisan news}: news that might not be entirely false, but they are phrased in a way that satisfies a specific political agenda and (d) \textit{satirical news}: any false content that has a humorous intent.

\subsection{Data Collection}
Reddit\footnote{https://www.reddit.com} is an inexpensive source of high-quality data \cite{Jamnik2017TheUO}. On Reddit, registered users tend to submit posts with richer content than Twitter, thus we are able to gather enough context for each user. Having enough users with rich contextual density is particularly beneficial for similarity assessment, which makes it the primary choice as the source for collecting disinformation spreaders and real news spreaders post histories. 

The data crawling was performed in a user-centric and iterative fashion.
To begin with, we manually compiled a list of 65 subreddits regarding controversial political  topics that were commonly discussed before the elections, such as general politics or the US presidential race, the SARS-CoV-2 pandemic, women's and men's rights, climate change, vaccines, abortion, gun control, 5G in general. For each of those subreddits, the most recent threads were crawled and inserted into a database. On this data, we performed the first iteration of the URL domain-based disinformation and real news spreader extraction to generate a list of Reddit user accounts with equal amounts of users for either class. We then collected the complete histories of all the users in said list, thus gathering all threads in which they participated in the list of political subreddits. All of those threads were inserted into the database from which, again, a now larger list of misinformation and real news spreaders can be extracted. This process was iterated until the dataset reached its current form. 

We show the subreddits included in the resulting dataset and the corresponding number of unlabeled, real and fake news posts they contain in Table~\ref{tab:subreddits}. In the parenthesis, we note the stance that each subreddit supports in its description. For each topic, the subreddits with a very low number of fake news posts, are grouped in the rows named ``Other''. In this table, the topics are shown based on a descending number of total fake news posts, the same stands for the subreddits that belong to them. For each topic, we opted for an equal distribution of political partisanship and stances, by selecting the same number of the most popular subreddits for each stance and for the same time period. 

As we can see, the largest portion of unlabeled, fake and real news posts are from the subreddit \textit{r/politics} which is a reddit with no specific political agenda for discussing the news regarding US politics. We can see that the conservative party seems to be posting more frequently based on the number of unlabeled posts. In addition, all topics have a skewed distribution of stances.

\begin{table}[t]
\small
 \setlength{\tabcolsep}{1pt}
\begin{tabular}{lrrr}
\toprule
\textbf{Subreddit}     & \textbf{\# unlabeled} & \textbf{\# real} & \textbf{\# fake} \\ \bottomrule
\multicolumn{4}{c}{General political debate}                                         \\ \hline

r/politics (no bias)	&2.399.254	&81.261	&3.869 \\
r/Conservative (right)	&346.042	&5.165	&2.784 \\
r/conservatives (right)	&24.310	&526	&453 \\
r/Republican (right)	&17.797	&500	&256 \\
r/ConservativesOnly (right)	&9.431	&57	&62 \\
r/democrats (left)	&11.747	&338	&41 \\
Other (mostly left) & 72.135 & 2.355  & 81    \\
\hline
\multicolumn{4}{c}{SARS-CoV-2}                                                       \\ \hline
r/NoNewNormal (anti)	&72.411	&1.941	&1.387 \\
r/LockdownSkepticism (no bias)	&62.480	&1.441	&275 \\
r/NoLockdownsNoMasks (anti)	&1.887	&82	&61 \\
r/Coronavirus (no bias)	&92.163	&2.753	&54 \\
Other (mostly no bias) & 21.697                & 606              & 53               \\ 
\hline
\multicolumn{4}{c}{Women's and men's rights}                                         \\ \hline
r/MensRights (men)	&57.654	&1.636	&501 \\
r/Egalitarianism (non-specific)	&83	&4	&42 \\
r/antifeminists (men)	&1.138	&44	&15 \\
Other (mostly women)                  & 1.399                 & 47               & 11               \\ 
\hline
\multicolumn{4}{c}{Climate change}                                                   \\ \hline
r/climateskeptics (questioning)	&38.606	&756	&856 \\
r/climatechange (science)	&7.858	&622	&153 \\
r/GlobalClimateChange (science)	&26	&2	&0 \\
r/climate (science)	&120	&12	&0 \\

\hline
\multicolumn{4}{c}{Vaccines}                                                         \\ \hline
r/DebateVaccines (no bias)	&32.635	&1.624	&637 \\
r/DebateVaccine (no bias)	&2.707	&57	&22 \\
r/TrueantiVaccination (anti)	&3.428	&48	&18 \\
Other (mixed anti and pro)  & 7.255  & 225    & 16    \\ 
\hline
\multicolumn{4}{c}{Abortion}                                                         \\ \hline
r/prolife (anti)	&7.109	&167	&82 \\
r/Abortiondebate (no bias)	&7.590	&84	&22 \\
Other (mostly pro)   & 5.228                 & 84               & 4                \\ 
\hline
\multicolumn{4}{c}{Guns}                                                             \\ \hline
r/progun (pro)	&10.774	&453	&61 \\
r/Firearms (pro)	&12.728	&200	&33 \\
r/GunsAreCool (pro)	&4.930	&233	&27 \\
r/gunpolitics (no bias)	&1.967	&61	&11 \\
r/guncontrol (anti)	&1.062	&206	&10 \\
Other (mostly pro)  & 9.744                 & 338              & 6                \\ 
\hline
\multicolumn{4}{c}{5G}                                                               \\ \hline
r/5GDebate (no bias)	&2.192	&19	&6 \\
\bottomrule
\end{tabular}
\caption{This table shows the names of the subreddits that belong to each topic and the corresponding number of unlabeled, real and fake news posts. The rows named ``Other'' contain the subreddits  with a low number of fake news posts for each topic.}
\label{tab:subreddits}
\end{table}

\begin{table}
\setlength{\extrarowheight}{2pt}
\begin{tabular}{r p{5.7cm}}
 \toprule
 \textbf{Date} & \textbf{Event Description} \\
 \bottomrule
\textbf{Feb 5}  & Trump is acquitted on the charges of abuse of power and obstruction of Congress.\\
 \hline
\textbf{Jul 11}  & Mail-in votes are encouraged. \\ 
 \hline
\textbf{Jul 30}  & Donald Trump threatens to postpone the election if it appears mail-in votes might go against him. (We regard this as if this had happened in August, since the effects of this political event would be still discussed during that month) \\
 \hline
\textbf{Aug 11}  & Joe Biden chooses Senator Kamala Harris (D-CA) as his running mate (event 1)\\
 \hline
\textbf{Nov 3}  & 2020 United States elections (event 2)\\
 \hline
\textbf{Jan 6}  & US Capitol is attacked by supporters of Trump (event 3) \\
 \hline
\textbf{Feb 24}  & Johnson \& Johnson's vaccine candidate receives emergency use authorization from the FDA (event 4) \\
 \bottomrule
 \end{tabular}
 \caption{Major political events coinciding with the peaks observed in the number of fake and real news posts from Figure~\ref{fig:num_labeled}}
 \label{tab:political_events}
\end{table}

\subsection{Media Domain Lists}

Likewise to the work of \newcite{baly2018predicting}, the website \emph{mediabiasfactcheck.com} was used as the main source for annotated news outlet domains. It was deemed a suitable resource for the study at hand as it offers annotations for two dimensions: the \emph{factuality level} and the \emph{political bias} of a large proportion of high frequented online news media.

Since we also opted for a binary label for the disinformation spreaders, we created a mapping for those labels. To be considered a disinformation domain, the \emph{mediabiasfactcheck} label has to be below or at \textit{Mixed} factuality level or labeled as satire, while the real news domains have to be at least \textit{Mostly factual} and between \textit{Right-Center} and \textit{Left-Center} political bias.

As for the credibility of the assigned annotations, the maintainers of \emph{mediabiasfactcheck.com} state that they \enquote{are looking at political bias, how factual the information is, and links to credible, verifiable sources} \cite{mbfcmethod}. In the description of their methodology, they also describe that they base the labels on reviews of at least 10 headlines and 5 news stories \cite{mbfcmethod}.

As a further resource to extend the list of disinformation media sources, an \enquote{index of fake-news, clickbait, and hate sites} \cite{cjrlist} by the \emph{Columbia Journalism Review}\footnote{https://www.cjr.org} was consulted. Its curators state that it was created by merging pre-existing fake news domain lists from various sources and then checking their actual invalidity with the fact checking platforms {PolitiFact} and {Snopes} \cite{cjrlist}. Finally, to ensure the quality of all annotations, we cross-matched the labels of the common domains by consulting both Snopes and Media Bias/Fact Check.

In total, in this way, we aggregated 1577 disinformation and 571 real news domains for our ground truth and post-level annotations.

\subsection{Binary Annotation.}
\label{sec:Binary_Annotation}
The users were annotated as \textit{misinformation spreaders} and \textit{real news spreaders} based on the posted web-links in their history. More precisely, we first extracted news links from the users’ posts using regular expression matching. To decide whether the extracted link was counted as {misinformation} or {real news}, its domain was matched with the two lists of domains of online news outlets, each corresponding to one class. Users were then labeled as \textit{misinformation spreaders} if they had at least two detected misinformation links in their post history, while for being \textit{real news spreaders} they had to have no shared links from the misinformation list and at least one link posted from the factual news list. 

\subsection{Fine-grained labels.} \label{sec:User_level_Annotation}

In addition to the binary separation of users into misinformation spreaders and real news spreaders, each user was annotated with the following factors by averaging over a float mapping of the labels from \emph{mediabiasfactcheck.com}, for a more fine-grained annotation.
\paragraph{Factuality degree (fd).} This factor represents the average level of factuality of each author, and is also in the range of [-3, +3] with each label corresponding to different scales;
 very low ($s_{vl} =-3 $), low ($s_{lf} =-2$), mixed ($s_{mx} =-1$), mostly factual ($s_{mf} =+1$), high ($s_{hf} = +2$), very high ($s_{vh} = +3$). Similarly, the factuality factor of each author is computed as follows:
 $$fd = \frac{ \sum_{\ell} s_{\ell} \cdot N_{\ell} }{  \sum_{\ell} N_{\ell}}$$ where $N_{\ell}$ in the number of posts labeled as $\ell \in [vl, lf, mx, mf, hf, vh]$
\paragraph{Political bias (pb).} This factor represents the level of partisanship and is a number in the range of [-3, +3] where each of the labels correspond to different scales ($s_{\ell}$);
 extreme left ($s_{el} = -3$), left ($s_{l} = -2$), center left ($s_{cl} = -1$), least biased ($s_{lb} = 0 $), center right ($s_{cr} = +1$), right ($s_{r} = +2$), and extreme right ($s_{er} =+3$). The political bias of each author is computed as:
 $$pb = \frac{ \sum_{\ell} s_{\ell} \cdot N_{\ell} }{  \sum_{\ell} N_{\ell}}$$ where $N_{\ell}$ in the number of posts labeled as $ \ell \in [el, l, cl, lb, cr, r, er]$
\paragraph{Science belief (sb).} This factor quantifies the level of belief in science and is a number in the range of [-1, +1] where each of the labels correspond to different scales ($s_{\ell}$); conspiracy theory article ($s_{c} =-1 $), science-based article ($s_{s} =1$). Similarly, the science factor of each author is computed as follows:
$$sb = \frac{ \sum_{\ell} s_{\ell} \cdot N_{\ell} }{  \sum_{\ell \in fl}}$$ where $N_{\ell}$ in the number of posts labeled as $\ell \in [s,c]$
\paragraph{Satire degree (sd).} This factor represents the level of satire in the fake news posts. The higher this factor is, the less intentional the misinformation spreading. It is in the range of [0, 1] and is computed as the number of satire posts $N_s$ divided by the number of fake news posts $N_{fn}$:
$$sd = \frac{N_s}{N_{fn}}$$

\paragraph{Discussion.}

Current datasets for fake news spreaders detection characterize a user as a fake news spreader based on whether they posted more than $n$ number of posts, which $n$ being an arbitrary number around two or three.
By introducing these fine-grained labels we pose some interesting questions to the research community. 
How many times should a user post about fake news in order to be considered as a fake news spreader? Should it also depend on what kind of fake news post they posted (e.g. a post from a pseudoscience source vs post from a source that has a mixed factuality reporting shouldn’t have the same gravity).
While satirical news is fake, the intent is usually humorous, however the dissemination of such news could be equally harmful. Should users who post from these sources also be considered as fake news spreaders? Should we consider a threshold of factuality degree instead of counting fake news posts to separate fake news posters and real news spreaders?

\subsection{Dataset Statistics} \label{sec:Dataset_Statistics}
The dataset comprises a total of 3.354.450 posts authored by 4,150 users with a class distribution of 74:26 of real news and fake news spreaders respectively, collected from January 2020 to April 2021. Misinformation spreaders had an average of 1240 posts, with this count being at 654 for the real news spreaders. In total, 2\% of the posts contained links to real news media, while 0.3\% pointed to domains from the misinformation list.

\begin{figure}[t]
    \centering
    \includegraphics[width=\linewidth]{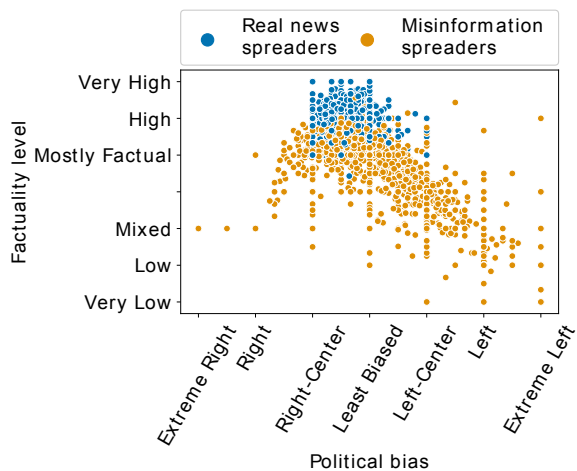}
    \caption{Factuality factor over political bias of each user.}
    \label{fig:ff_bias}
\end{figure}

Using the post-level annotations from Section~\ref{sec:User_level_Annotation}, the political biases of the users can be looked at:
41.17\% of the users that have left wing political bias are misinformation spreaders, while 58.82\% of them are real news spreaders.
91.58\% of the users that have right wing political bias are fake news spreaders, while only 8.41\% of them are real news spreaders.
Figure~\ref{fig:ff_bias} depicts the factuality factor over political bias of each user.
While there is an apparent correlation (Pearson correlation of -0.45) between the political bias and factuality of the users, it is important to note that this effect is not an isolated case or a problem that rises from the process of collecting our data, in fact, this phenomenon has been observed by many researchers \cite{10.1145/3341161.3343536} who show that there is indeed a high correlation between the perceived bias of a publisher and the trustworthiness of news content. In addition, \cite{doi:10.1126/sciadv.abf1234} showed that  US conservatives are uniquely susceptible to misinformation regarding the political events and generally  political extremes (both the left and the right) are substantially susceptible to conspiracy beliefs. Note that from Table~\ref{tab:subreddits}, we can see a higher posting activity from the right wing party compared to the left wing, which leads us to the conclusion that right-wing supporters might be more active in social platforms compared to left-wing supporters.

\begin{figure}[t]
    \centering
    \includegraphics[width=0.8\linewidth]{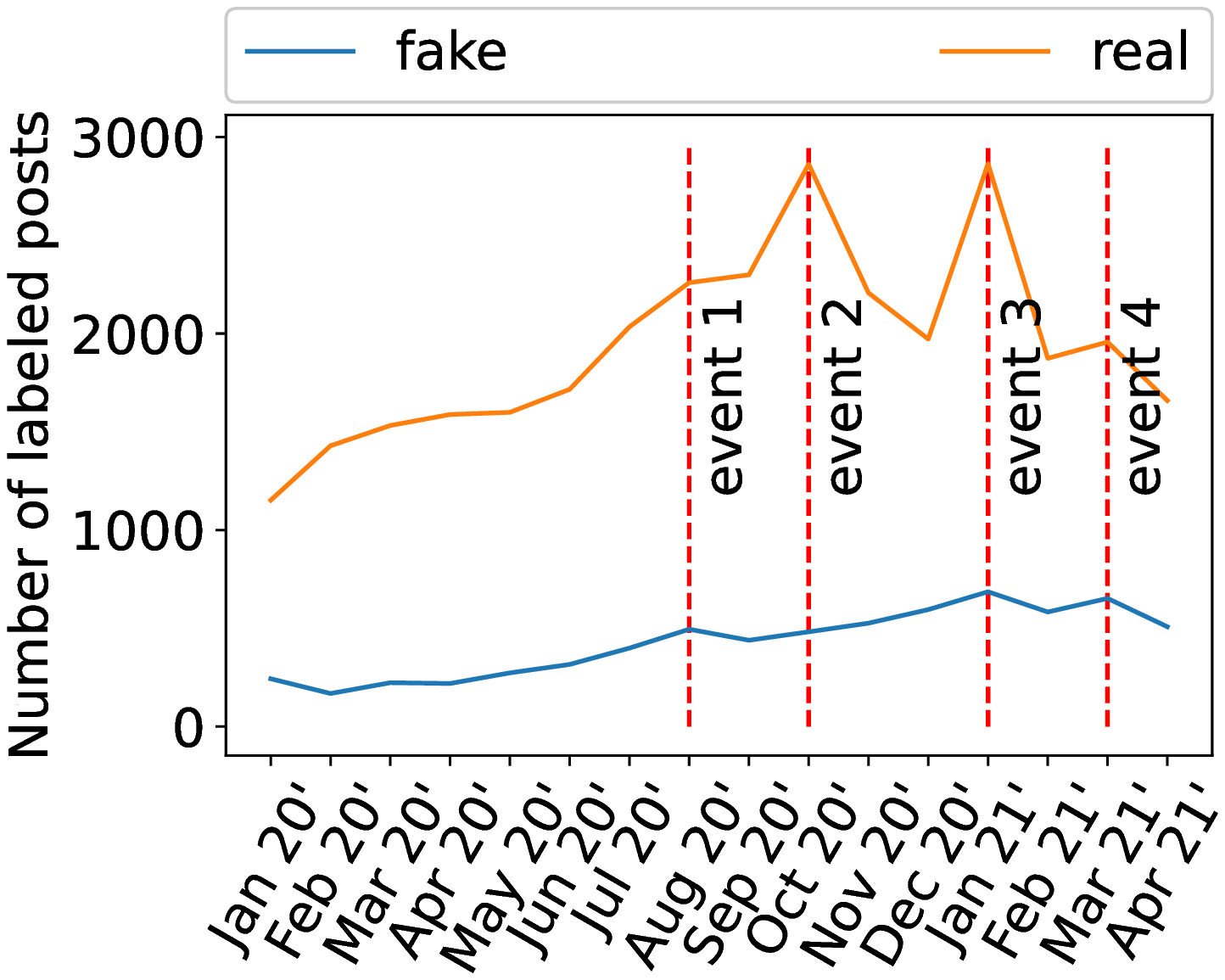}
    \caption{Number of fake news posts and real news posts associated with the political events from Table~\ref{tab:political_events}}
    \label{fig:num_labeled}
\end{figure}

The timestamps and thread structure of all stored posts is preserved in the dataset, in order to encourage a more comprehensive analysis of the users and their posting behavior. Figure~\ref{fig:num_labeled} shows the number of fake news and real news posted per month. 
We also provide a list of pivotal political events\footnote{{https://en.wikipedia.org/wiki/2020\_in\_United\_States \_politics\_and\_government},\\ {https://en.wikipedia.org/wiki/2021\_in\_the\_United\_States}} that happened during this time period in Table~\ref{tab:political_events}. We can see that these events coincide with the increase in the number of fake news and real news posts. We can see an obvious increase in real news right until the US elections and a sudden increase during the attack on the Capitol. This is logical since the elections were scheduled and discussed months before they happened while the attack was an event that developed over a few days. A smoother curve is observed for the fake news, where the numbers do seem to fluctuate in the same manner during these events, but not to the same degree.

\section{Encoding the Users} \label{sec:Encoding_Users}

\subsection{Problem Formulation}
We denote the user to be classified as  ${u}^i \in \mathcal{U} = \{ {u}^1, {u}^2, \dots, {u}^N \}$. Each user ${u}^i $ is associated with a posting history 
$\mathcal{H}^{i} =\{(p_{1}^{i}, t_{1}^{i}), (p_{2}^{i},t_{2}^{i}),\dots, (p_{L^i}^{i},t_{L^i}^{i})\} $ where $p_{k}^{i}$ is a text authored by the user $u^i$, posted at time $t_{k}^{i}$ where $t_{1}^{i} < t_{2}^{i} < \dots < t_{L^i}^{i} $  and $L^i$ is the individual posting history length of each user $u^i$. 
\paragraph{Fake news spreader detection.}
For the following experiments we utilize the binary labels introduced in Section~\ref{sec:Binary_Annotation}. We therefore formulate the author profiling problem as a binary classification task to predict the class $y^i$ of the user, where $y^i \in  \{$misinformation spreader, real news spreader$\}$. 

\paragraph{Political bias identification.}
We utilize the fine-grained labels of the political bias introduced in Section~\ref{sec:User_level_Annotation}. The left-wing supporters are the users with $pb<-0.5$, while the right-wing supporters are those with $pb>0.5$. Accordingly, the identification of partisanship is defined as a binary classification task to predict the class $y^i$ of the user, where $y^i \in  \{$left wing, right wing$\}$. 

\subsection{User representations}
\label{sec:user_emb}
\paragraph{BERT-based representations}
We use Sentence-BERT (SBERT) \cite{reimers-gurevych-2019-sentence} to obtain the embedding $e_{k}^{i}$ of each user’s individual historical posts $p_{k}^{i}$. SBERT is a modification of BERT that is specifically designed to produce semantically meaningful sentence embeddings, and has achieved state-of-the-art performance on various challenging datasets 
\cite{10.5555/2387636.2387697,Cer_2017,marelli-etal-2014-sick}, rendering this encoding method particularly suitable for representing the users.

We want to encode the users' historical context $\mathcal H^{i}$ by obtaining their user representations $E^i \in \mathbb{R}^{d_b}$. \newcite{Lee2020CapturingWO} empirically showed that simple average sentence embeddings compare favorably to more complex methods. Each user's historical encoding is averaged over the individual posting history length of each user ${L^i}$, as:
$$
{E^i} = \frac{1}{L^i} \sum_{k=1}^{L^i} e_{k}^{i}
$$

\paragraph{User2Vec} 
In addition, we also adopt User2Vec \cite{amir-etal-2016-modelling} to compute the initial user representation. $E^i \in \mathbb{R}^{d_u}$ of user $u^{i}$ based on their corresponding historical posts $\mathcal{H}^{i}$, optimizing the conditional probability of texts given the author.
 
\paragraph{Encoding the psycho-linguistic features}
In order to analyze the relationship between users' tendency to spread fake news and their personality traits and mental processes, we use the Big Five Model and LIWC software respectively. The two methodologies are described hereafter.

The Big Five Model (BFM) \cite{FFM} assumes that human personality can be summarized in five main aspects: \textit{(i) openness to experience}, \textit{(ii) conscientiousness}, i.e. the interactions between rational thought and instincts, \textit{(iii) agreeableness}, or the intensity of individuals' reactions within the social context, \textit{(iv) extraversion}, and \textit{(v) emotional instability}. After defining these basic dimensions, this approach argues for the existence of semantic associations between them and specific sets of adjectives which are recurrent in the natural language when describing individuals' psychological traits. Accordingly, \newcite{NC} derive a personality score with the following process: for each factor, they compute the mean of all the cosine similarities between the embedding representations\footnote{The word embeddings are produced by a Word2Vec architecture, pre-trained on the Google News Corpus.} of every word in the input text and every benchmark adjective empirically observed as to be able to encode that precise personality trait; the higher this average similarity, the greater the evidence of a given factor. Neuman and Cohen also included 9 extra factors describing mental disorders, like \textit{paranoia}, and \textit{narcissism}.

The Linguistic Inquiry and Word Count software (LIWC) \cite{LIWC_bib} applies a lexicon-based method for mapping the text into 64 psycho-linguistic categories defined to obtain evidences of many mental processes underlying the natural language, and grouped into 2 macro-categories: \textit{(i) linguistic dimensions}, i.e. function words, common verbs and adjectives, etc. and \textit{(ii) psychological processes} of many kinds, including the affective, cognitive, and social type. In conclusion, the LIWC representation of one document consists in the set of relative frequencies for the categories, according to the number of words identified in the text that are associated with each of them. 
Again, both psycho-linguistic encodings are achieved by an averaging operation over the post-level ones.
In particular, it was preferred to extract the values of the LIWC features as means of the relative frequencies at the post level in order to extract the average incidence of each category within the single publication, with the aim of avoiding that the calculations were biased towards the most frequent classes within the composition of the global user discourse. 

\section{Methodology} \label{sec:methodology}

\subsection{Graph construction} \label{sec:Graph_construction}

\begin{figure}[t]
    \centering
    \includegraphics[width=0.9\linewidth]{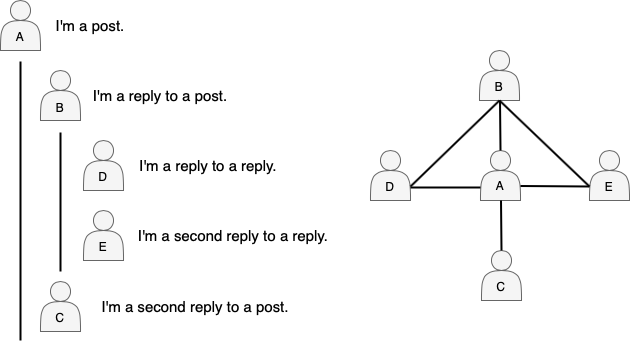}
    \caption{Transforming a post/reply tree in social media into a social graph network.}
    \label{fig:users_interactions}
\end{figure}

Social science argues that like-minded people tend to interact more with each other \cite{Bahns17}, therefore we construct the social graph in a way that captures the users' social interactions with each other. We define as social interaction the replies and mentions in a post thread. For each thread of posts, we connect all the chain of replies to the root (i.e. the original post) of the conversation and all mentions/replies to each other. Following, these connections are translated to user connections in the social graph. This method is more clearly depicted in Figure~\ref{fig:users_interactions}. The social graph $\mathcal{G} = (\mathcal{V}, \mathcal{E})$ is comprised by a set of user nodes $\mathcal{V}$ and a set of edges $\mathcal{E}$ between these users. 

\subsection{Graph Attention Network}

Graph neural networks are able to leverage the semantic and social relations between users \cite{Wu_2021}. As users have a different influence on one another, we need to focus on users that have more relevant connections with higher influence. To model the gravity of the influences of the neighbourhood to a node, we use Graph Attention Networks (GAT) \cite{velickovic2018graph}. GAT attends to the neighborhood of each user and  assigns an importance score to the connections that contribute more to the detection of misinformation spreaders. The input to a GAT layer is a set of users embeddings $\mathcal{E}= \{E^{1},\dots,E^{N}\}$  where $N = |\mathcal{U}|$. A GAT layer produces updated features, $\widetilde{ \mathcal{E}} = \{\widetilde{E^{1}}, \dots, \widetilde{E^{N}}\}$, where $\widetilde{E^{i}} \in \mathbb{R}^{d_g}$. First, the GAT layer applies a shared linear transformation by a weight matrix $ \mathbf{W} \in \mathbb{R}^{d_g \times d_b}$. Then, we apply a shared self-attention mechanism to each node $i$, using the neighbourhood $\mathcal{N}(i)$. The normalized attention weight $\alpha_{ij}$ between node $i$ and neighbour node $j$ is computed as follows:
\begin{equation}
    \alpha_{ij} = \frac{exp(LeakyReLU(a_w^{\intercal}\,[\mathbf{W}\, E^{i} \; \| \; \mathbf{W} \, E^{j}] )}{\sum_{k \in \mathcal{N}(i)} exp(LeakyReLU(a_w^{\intercal} \, [\mathbf{W} \, E^{i} \; \| \; \mathbf{W} \, E^{k}] )}
\end{equation}
\noindent where $\top$ represents the transpose and $\|$ is the concatenation operation. $a_w \in \mathbb{R}^{2d_g}$, is a trainable parameter vector. The attention weights  $\alpha_{ij}$ represent the importance of relation from node $i$ to node $j$. To stabilize the learning process, we employ a multi-head attention \cite{vaswani2017attention}. We compute the output representation of a node $\widetilde{E^{i}}$ as follows: 
\begin{equation}
     \widetilde{E^{i}} =  \text{ReLU} \left(\frac{1}{K} \sum_{k=1}^{K} \sum_{j \in \mathcal{N}(i)} \alpha_{ij}^{k}  \mathbf{W}^k E^{j} \right)
\end{equation}
\noindent where, $\mathbf{W}^k$ denotes normalized attention weight and linear transformation for $k$-th head.

\paragraph{Classification Layer} The overall learned representations for each user, are forwarded into a linear layer parameterized by a weight matrix $\mathbf{W^{o}} \, \in \mathbb{R}^{d_o \times d_r}$. The final prediction is computed as:
\begin{equation}
    \hat{y} = softmax(\mathbf{W^{o}} \, \Gamma(\overline{h})).
\end{equation}
Given the true label $y$ for a user, we use cross-entropy loss to calculate the loss $L$ as follows:
\begin{equation}
    L = - \sum_{i = 1}^{N} y_i \, ln(\hat{y_i}) + (1 - y_i) \, ln(1 - \hat{y_i}).
\end{equation}

\section{Experiments}

\subsection{Models used}

We compare our graph-based model as described in Section~\ref{sec:methodology}, with a Support Vector Machine (SVM), Logistic Regression (LogReg), and a Random Forest (RnFor) classifier which are trained by using the following features:

\noindent\textbf{UBERT}: We use the SBERT embeddings of the documents averaged over the user's history as feature vectors, as described in Section~\ref{sec:user_emb}.

\noindent\textbf{User2Vec}: To initialize the user feature vectors, we use User2Vec over the vocabulary of each user during their history. 

\noindent\textbf{Psycholing}: We concatenate both LIWC and BFM features, to compute an initial feature vector for the users.

\subsection{Performance evaluation and ablation study}

Table~\ref{tab:gat_results} shows GAT's $F_1$ score on the Reddit dataset for the fake news spreader detection task. We compare the graph-based results by using different initialization methods, namely UBERT, User2Vec, psycho-linguistic, concatenation of User2Vec and psycho-linguistic features, and random vectors.  Interestingly, the proposed model achieves the best performance by utilizing User2Vec, despite having lower dimensionality than UBERT. This is mainly attributed to the fact that User2Vec embeddings were obtained based based on this dataset, while UBERT was pre-trained on a general-use corpus. The psycho-linguistic features, on their own, perform rather poorly with GAT and  concatenating them to User2Vec does not contribute to the performance. However, the psycho-linguistic features perform comparably to UBERT in the non-neural baselines, which is in line with the observations of \newcite{Rashkin2017Truth}. 
\begin{table}[ht]
\centering

\begin{tabular}{l|c} 

\multicolumn{2}{c}{\textbf{Fake News Spreader Detection}} \\
\textbf{Model}& \textbf{$F_1$ score}  \\ 
\hline
GAT + User2Vec (200)                                       & \textbf{61.6\%}  \\
GAT + UBERT (768)                                          & 61.2\%           \\
GAT + Psycholing (83)                                      & 53.6\%           \\
GAT + User2Vec + Psycholing (283)                          & 59.4\%           \\
GAT + Random (200)                                         & 47.8\%           \\

\end{tabular}
\caption{Comparison of different user embeddings techniques for the GAT model on the fake news spreader detection task. Reported values are the $F_1$-scores over a 5-fold Cross Validation. Bold denotes the best overall performance on the task.}
\label{tab:gat_results}
\end{table}

Table~\ref{tab:baselines} shows the $F_1$ score of the baseline models for both the the political bias and fake news spreader detection tasks. For the political bias identification task, UBERT consistently obtains better results than User2Vec, and achieves the best result with SVM.
On the other hand, for the Fake news spreader detection task, we observe the reversed behavior. User2Vec consistently obtains significantly better results than UBERT, and achieves the best result with a Random Forest classifier.

\begin{table}[ht]
\small
\begin{tabular}{l|cc|cc} 

                                    & \multicolumn{2}{c|}{\textbf{Political Bias}} & \multicolumn{2}{c}{\textbf{Fake News Spreader}}  \\
{\textbf{Model}} & \textbf{UBERT}   & \textbf{User2Vec}         & \textbf{UBERT} & \textbf{User2Vec}               \\ 
\hline
SVM                                 & \textbf{66.2\% } & 63.0\%                    & 53.9\%         & 61.1\%                          \\
LogReg                              & \ul{64.7}\%   & 62.8\%                    & \ul{58.6}\% & 59.8\%                          \\
RnFor                               & 64.9\%           & \ul{63.5}\%            & 49.7\%         & \textbf{\ul{61.3}}\%         \\

\end{tabular}

\caption{Comparison of different user embeddings
techniques for the baseline models for both political bias and fake news spreaders detection. Reported values are the $F_1$-scores over a 5-fold Cross Validation. Bold denotes the best overall performance on the task.}
\label{tab:baselines}
\end{table}

Table~\ref{tab:psyablation} shows the ablative results of the psycho-linguistic features on the Reddit dataset for both political bias and fake news spreaders detection. In general, psycho-linguistic features show a significantly higher effectiveness in distinguishing users on the basis of political bias. Detected mental processes appear to be significantly more useful than personality factors: this result is coherent with the study conducted through the LIWC software by \newcite{PennebekerPolitics} about the link between political ideology and language use. Most relevant mental process is the \textit{affective} kind, which correlates negatively with the target class, suggesting that right-biased users tend to express fewer emotions such as anxiety, anger and sadness in the text. As regards the other task, the BFM encoding appears slightly more effective for identifying fake news spreaders. Indeed, since personality regulates the behavior in real contexts, it is reasonable to assume it to be also influential within virtual communities. The dominant factor is here the \textit{openness to experience}: as expected, in those who spread fake news, there is greater rejection or less curiosity towards ideas outside their belief system. Also, the \textit{schizotypy} disorder appears relevant, consistent with previous empirical observations \cite{Buckels2018}. 

We note that the psycho-linguistic features are not adaptive to the tasks since they are lexicon-based, therefore the embedding-based features achieve significantly higher $F_1$ scores in the political bias detection task. By comparing all results for the fake news spreader detection task, we observe that the GAT model outperforms all baselines. Therefore, the social interactions constitute a promising tool for predicting the behavior of unseen users. 

\begin{table}[ht]
\small
\setlength\tabcolsep{3.5pt}
\begin{tabular}{l|c c c  | c c c }

& \multicolumn{3}{c|}{\textbf{Political Bias}}        & \multicolumn{3}{c}{\textbf{Fake News Spreader}}\\
\multicolumn{1}{c|}{\textbf{Model}}        & \multicolumn{1}{c}{\textbf{LIWC}} &  \multicolumn{1}{c}{\textbf{BFM}} & \multicolumn{1}{c|}{\textbf{Both}} & \multicolumn{1}{c}{\textbf{LIWC}} &  \multicolumn{1}{c}{\textbf{BFM}} & \multicolumn{1}{c}{\textbf{Both}} \\ \hline

 {SVM}       &55.1\%   &38.8\%  & 61.0\%     &56.2\%	&51.0\%	&53.9\%
                            \\
 {LogReg}    &\underline{63.6}\%    &51.5\%    & \underline{\textbf{63.9}}\%    &\underline{58.3}\%	&55.1\%	&\underline{58.3}\%
                                \\
 {RnFor} &56.6\% &\underline{54.8}\% & 61.7\%       &55.9\%	&\underline{\textbf{58.4}}\%	&54.8\%                              \\

\end{tabular}
\caption{Ablation study over the psycho-linguistic features and their combination for both political bias and fake news spreaders detection. Reported values are the average $F_1$-scores over a 5-fold Cross Validation. Underlines denote the best result for the combination of features considered, while bold denotes the best overall performance on the task. 'Both' indicates the concatenation of both representations.}
\label{tab:psyablation}
\end{table}

\section{Conclusion}
In this study we introduce a new user-centered dataset for misinformation spreader analysis, monitoring political discussions on Reddit since the beginning of 2020.
We create a dataset that contains over 4K users with 3.4M Reddit posts, covering the time period before and after the US presidential elections. Apart from the fake news/real news distinction, the dataset contains fine-grained labels about the users' credibility level and political bias. As far as we are aware, this is the first fake news spreader dataset that simultaneously captures both the long-term context of user's historical posts and the interactions between users. To create the first benchmark on our data, we provide methods for identifying misinformation spreaders by utilizing the social connections between the users along with their psycho-linguistic features. In a subsequent analysis we observe that social connections increase robustness over content features, that detecting affective mental processes correlates negatively with right-biased users, and that the openness to experience factor is lower for those who spread fake news.

\section{Ethical Considerations and Limitations}
The ability to automatically approximate personal characteristics of online users in order to improve natural language classification algorithms requires us to consider a range of ethical concerns. Researchers are advised to take account of users’ expectations \cite{shilton2016we,townsend2016social} when collecting public data such as Reddit. All user data is kept separately on protected servers, linked to the raw text and network data only through anonymous IDs.
In addition, any user-augmented classification efforts risk invoking harmful stereotyping, as the algorithm labels people as misinformation spreaders. These can be emphasized by the semblance of objectivity created by the use of a computer algorithm \cite{koolen-van-cranenburgh-2017-stereotypes}.

\section*{Acknowledgements}
This work has been supported by the German Federal Ministry of Education and Research (BMBF) as a part of the Junior AI Scientists program under the reference 01-S20060. The work at the Universitat Politècnica de València was in the framework of XAI-DisInfodemics: eXplainable AI for disinformation and conspiracy detection during infodemics (PLEC2021-007681) funded by the Spanish Ministry of Science and Innovation, and IBERIFIER: Iberian digital media research and fact-checking hub (INEA/CEF/ICT/A202072381931, n. 2020-EU-IA-0252) funded by the European Digital Media Observatory. 

\section{Bibliographical References}\label{reference}

\end{document}